\documentclass[aps,prb,reprint,showpacs]{revtex4-1}
\usepackage[pdftex]{graphicx}  
\usepackage{dcolumn}   
\usepackage{bm}        
\usepackage{amssymb}   
\usepackage{amsmath}
\usepackage{hyperref}
\usepackage[latin1]{inputenc}

\begin{document}

\title{Adiabatic and local approximations for the Kohn-Sham potential in time-dependent Hubbard chains}
\date{\today}

\author{L. Mancini}
\affiliation{Department of Physics, University of York, Heslington, York YO10 5DD, United Kingdom}
\affiliation{University of Bologna, Department of Physics and Astronomy, viale Berti Pichat 6/2 40127, Bologna, Italy}
\altaffiliation{Current address: Groupe de Physique des Mat\'eriaux, UMR CNRS 6634, University of Rouen, 76801 St. Etienne du Rouvray, France}
\author{J. D. Ramsden}
\affiliation{Department of Physics, University of York, Heslington, York YO10 5DD, United Kingdom}
\author{M. J. P. Hodgson}
\affiliation{Department of Physics, University of York, Heslington, York YO10 5DD, United Kingdom}
\author{R. W. Godby}
\affiliation{Department of Physics, University of York, Heslington, York YO10 5DD, United Kingdom}

\begin{abstract}
We obtain the exact Kohn-Sham potentials $V_{\mathrm{KS}}$ of time-dependent density-functional theory for 1D Hubbard chains, driven by a d.c.\ external field, using the time-dependent electron density and current density obtained from exact many-body time-evolution. The exact $V_{\mathrm{xc}}$ is compared to the adiabatically-exact $V_{\mathrm{xc}}^{\mathrm{ad}}$ and the ``instantaneous ground state'' $V_{\mathrm{xc}}^{\mathrm{igs}}$. The latter is shown to work effectively in some cases when the former fails. Approximations for the exchange-correlation potential $V_{\mathrm{xc}}$ and its gradient, based on the local density and on the local current density, are also considered and both physical quantities are observed to be far outside the reach of any possible local approximation. Insight into the respective roles of ground-state and excited-state correlation in the time-dependent system, as reflected in the potentials, is provided by the pair correlation function. 
\end{abstract}
\maketitle

\section{Introduction}
With the prospect of technologies based on molecular devices, and the necessity of being able to describe systems in which devices interact with a time-dependent (TD) environment, there is a need for \textit{ab initio} methods to realistically describe correlated electronic systems subjected to TD external fields. Density-functional theory~\cite{PhysRev.136.B864, PhysRev.140.A1133} has proven to be an effective tool for describing ground-state properties of a great variety of different systems, often based on local approximations such as the local-density approximation (LDA)~\cite{RevModPhys.61.689}. In the time-dependent generalization of DFT, TDDFT~\cite{PhysRevLett.52.997}, the LDA is often used adiabatically (ALDA)~\cite{PhysRevLett.76.1212}.

Several studies have shown that the ALDA is able to describe properly only a limited range of time-dependent physical systems and that it frequently breaks down~\cite{gritsenko:8478, RevModPhys.74.601, PhysRevB.49.1849, PhysRevB.61.13431}. There is a need for more reliable approximations for the exchange-correlation potential in time-dependent correlated electron systems. In this paper we consider the Hubbard model~\cite{Hubbard26111963}, in which the electron-electron interaction takes a simple short-range form, aiming to obtain insight which may be extended to more general correlated electron systems.  The simplicity of the model means that reliable results may be obtained for substantial numbers of interacting electrons with reasonable computational effort.

In this paper we study 1D atomic chains represented by the Hubbard model, subjected to external TD perturbing potentials, for which a Runge-Gross theorem has been proved~\cite{PhysRevB.86.125130}. Using the electron density and the current density obtained from the exact time evolution of the interacting system, we deduce the TDDFT Kohn-Sham (KS) potential which causes the auxiliary KS system of non-interacting electrons to reproduce the TD electron density of the interacting system, and in particular its exchange-correlation part $\emph{V}_{\mathrm{xc}}$. 
One-dimensional Hubbard chains, owing to their relative simplicity in comparison to other many-electron systems, have provided a fruitful environment for the exploration of exact TDDFT in recent years (e.g. Ref. \onlinecite{PhysRevLett.101.166401}).

\section{Time-dependent Hubbard chains and their TDDFT description}

The systems we consider are described by the Hamiltonian
\begin{equation*}
\label{eq:HubHam}
\widehat{H} = U \sum_{R=0}^{N-1} \widehat{n}_{R\uparrow} \widehat{n}_{R\downarrow} - t \sum_{ \langle RR' \rangle \sigma } \widehat{\Psi}_{ R\sigma }^\dagger \widehat{\Psi}_{R'\sigma}^{ }  + \sum_{R\sigma} V_{R}^{\mathrm{ext}}(\tau) \widehat{n}_{R \sigma} 
\end{equation*}
where the on-site energy of an electron has been set to zero, $ U $ is the interaction energy, $ t $ is the hopping parameter and $ V_{R}^{\mathrm{ext}}(\tau) $ is a local external TD field  \footnote{We use $ \hbar = 1 $ and $ t=1 $, thus the unit of time is $ \hbar /t $.}. The notation $\langle RR' \rangle$ denotes nearest-neighbor sites, $\widehat{n}_{R \sigma} = \widehat{\Psi}_{ R\sigma }^\dagger \widehat{\Psi}_{R\sigma}$, and $\sigma = +1,-1$ denotes the spin. The system is prepared in its ground state. The external perturbing potential $ V_{R}^{\mathrm{ext}}(\tau) $ decreases with constant gradient from the right-most site to the left-most site (on which its value is zero), representing a uniform d.c.\ electric field that tends to drive electrons to the left. We study two different perturbations to this system: (a) the field is turned on at its full strength $V_{max}$ at $\tau=0$, remaining constant thereafter (`sudden'); (b) the same field is turned on gradually as a linear function of time from $\tau=0$ to $\tau_0$, remaining constant thereafter (`slow'). We consider chains whose number of sites $ N $ ranges from 2 to 8, in the half-filling configuration (i.e. with $N$ electrons) -- in which the chains exhibit the characteristics of a Mott-Hubbard insulator, with strongly localized electrons -- and also the quarter-filling ($N/2$ electrons) configuration.

\begin{figure}[h]
\begin{center}
\includegraphics[scale=0.34]{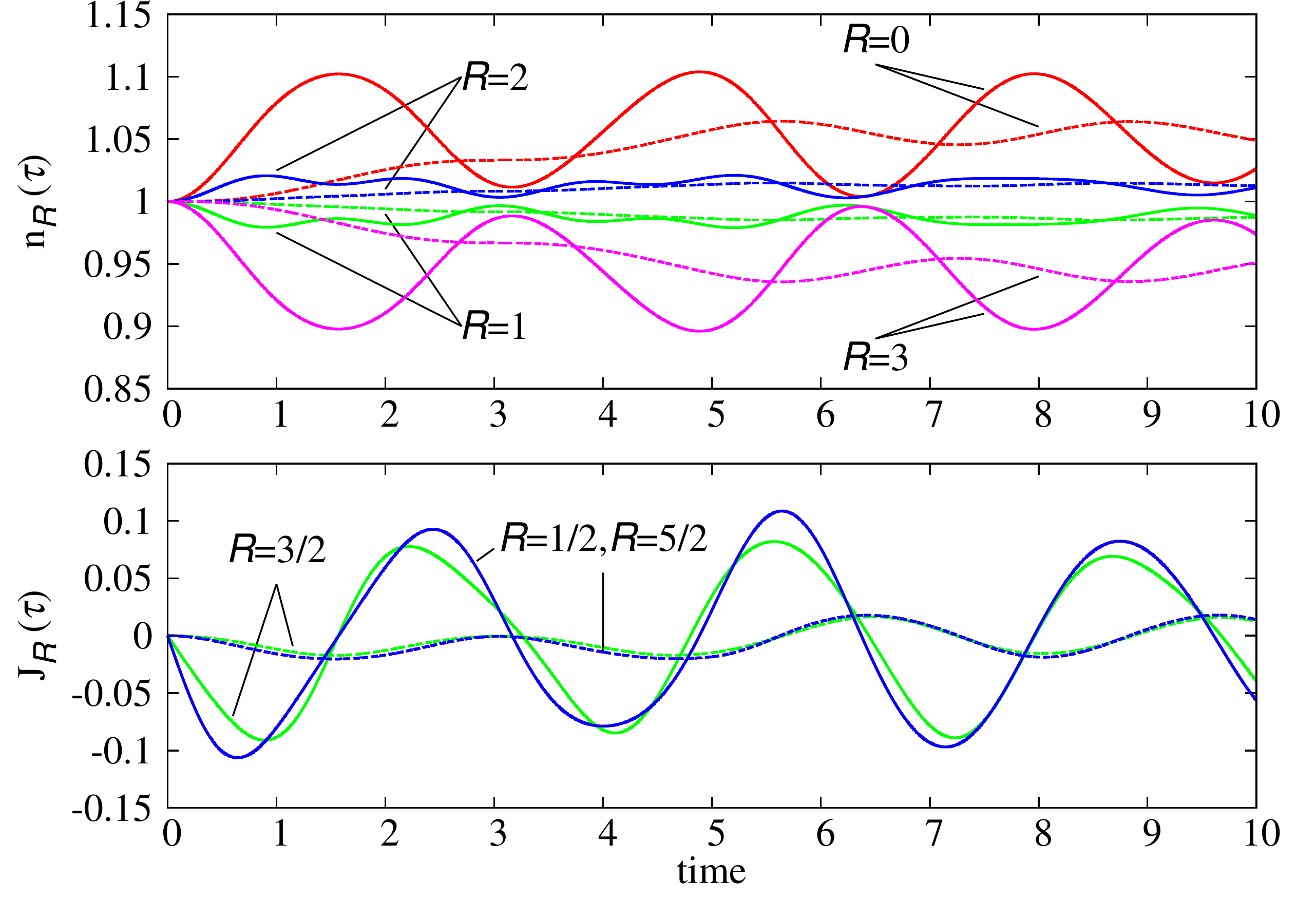}
\caption[density]{(color online) Local charge $ n_{R}(\tau) $ and current $ J_{R+1/2}(\tau)$ density  for a chain with four sites and four electrons, with $ U=2 $, $ t=1 $ and $ V_{\mathrm{max}}=0.5 $. Full lines are for the sudden potential, and dashed lines are for a slow potential with $\tau_0=5$. In this 4-site half-filled chain the charge density variation is symmetric and $J_{1/2}=J_{5/2}$.}

\label{fig:density}
\end{center}
\end{figure}

The interacting system is prepared in its ground state at $\tau=0$ using exact diagonalization of the Hamiltonian. The exact time propagation of the many-body (MB) wavefunction thereafter is computed from the time-dependent Schr\"{o}dinger equation using the Crank-Nicolson algorithm~\cite{CrankNicolson}. From $\vert \Psi (\tau) \rangle $ the local TD electron densities $ n_{R}(\tau) $ for each site are obtained, and the current $ J_{R+1/2}(\tau) $ between each pair of neighboring sites is defined as $ J_{R+1/2}(\tau) = J_{R-1/2}(\tau) + [\partial n_{R}(\tau)/\partial \tau] $, where the current densities beyond the end sites, $J_{-1/2}$ and, by implication, $J_{N-1/2}$, are zero. In Figure ~\ref{fig:density} we show the charge $ n_{R}(\tau) $ and the current $ J_{R+1/2}(\tau)$ densities in a 4-site half-filled chain, for the two different time-dependent potentials, with $ U=2 $, $ t=1 $ and $ V_{\mathrm{max}}=0.5 $.
In the half-filling configuration, the ground-state density of the interacting system is unity on each site. When the perturbing potential is applied, the charge density increases or decreases from this value depending on the site. Owing to the symmetry of the system, the charge density variation is, at all times, observed to be an odd function relative to the center of the chain. The inter-site current density oscillates around zero and is, correspondingly, an even function. Away from half filling, the ground-state charge density varies from site to site while retaining its mirror symmetry, while under the subsequent time-evolution the symmetry is broken.
\begin{figure}[h]
\begin{center}
\includegraphics[scale=0.34]{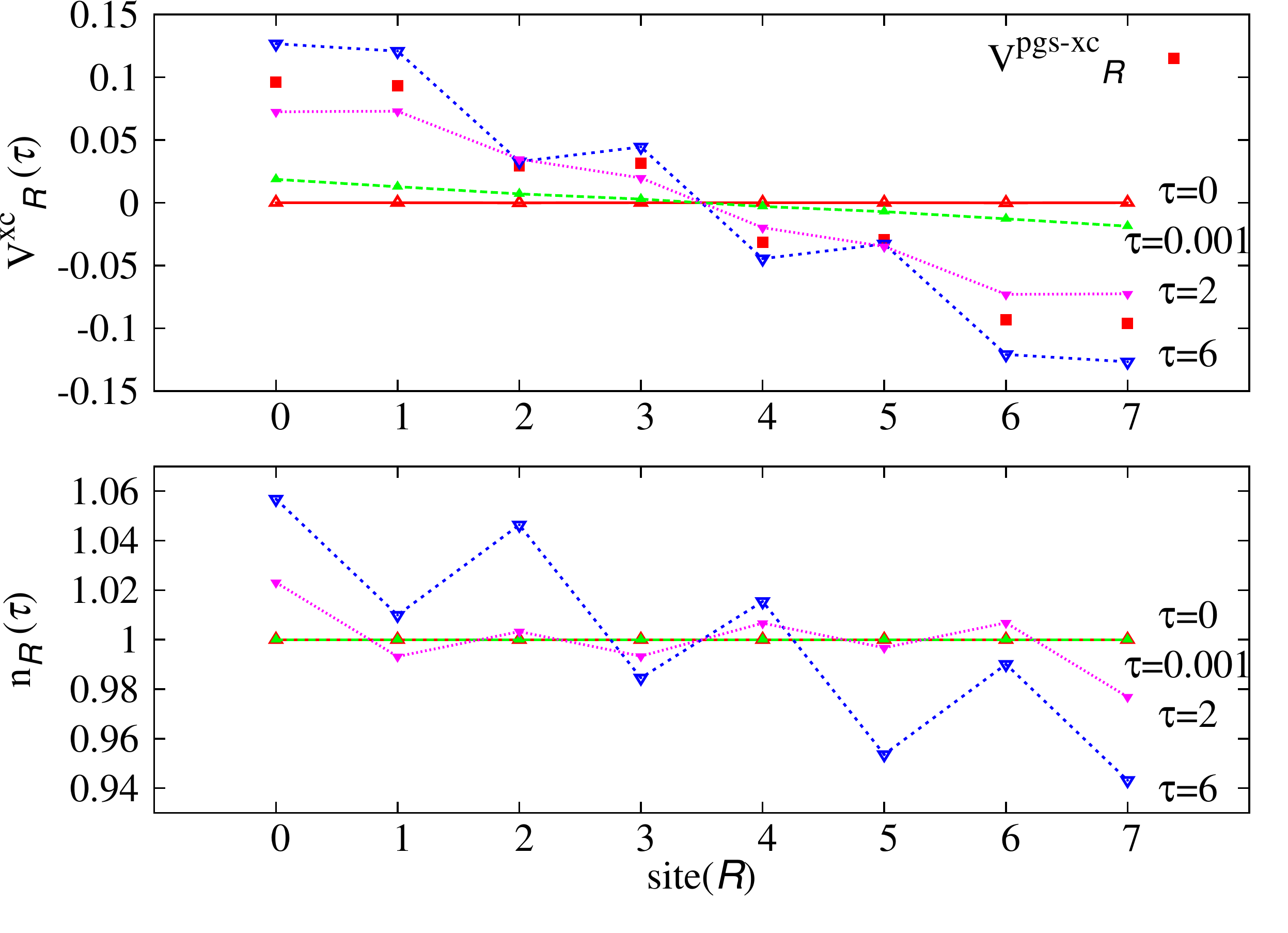}
\caption[Vn]{(color online) Exchange-correlation potential and charge density as a function of the position for different times $\tau$, for a half-filled 8-site chain subjected to an sudden $ V_{R}^{\mathrm{ext}}(\tau) $. When $ V_{R}^{\mathrm{ext}} $ is applied, $ V_{R}^{\mathrm{xc}} $ acquires a non-zero slope in the first time-step, even though the charge density is essentially indistinguishable from that of the ground state $ n_{R}(0) $. At later times, $ V_{R}^{\mathrm{xc}} $ slightly varies around the ground-state exchange-correlation potential of the perturbed Hamiltonian $ V_{R}^{\mathrm{pgs-xc}} $.}

\label{fig:vxc}
\end{center}
\end{figure}

The KS potential $ V_{R}^{\mathrm{KS}} $, which enables the KS system to reproduce the exact charge and current densities, is found via Powell's conjugate direction method~\cite{Powell01011964} employed to minimize the quantity \footnote{We weight charge and current differences equally because this choice produces good numerical stability.}
\begin{equation*}
\sum_{R \sigma}[(n_{R \sigma}^{\mathrm{KS}}(\tau) -  n_{R \sigma}(\tau))^2 + (\dot{n}_{R \sigma}^{\mathrm{KS}} - \dot{n}_{R  \sigma})^2].
\end{equation*}
$ V_{R}^{\mathrm{KS}} $ is the sum of the external potential $ V_{R}^{\mathrm{ext}} $ and the Hartree-exchange-correlation potential $ V_{R}^{\mathrm{Hxc}} $.  This is initially determined for the ground state (with no current term necessary), and subsequently the KS potential is determined at each time-step so that the interacting charge and current densities continue to be reproduced. (The potential $ V_{R}^{\mathrm{Hxc}} $ is defined up to an additive overall $R$-independent constant by this requirement, so we choose its value to be always zero on the left-most site in all systems studied.) The resulting KS electron densities are always within $10^{-6}$ of the exact values, for the strengths of the external field $ V_{R}^{\mathrm{ext}} $ considered in this paper \footnote{For high values of $ V_{R}^{\mathrm{ext}} $ the Kohn-Sham (KS) system appears to be numerically unstable. A similar inability to determine the KS potential, for large perturbing fields, has previously been observed for a short-range interaction lattice impurity model~\cite{schmitteckert2013exact}, where it was suggested to be a violation of (non-interacting) $v$-representability, which has been shown to be an issue for lattice systems~\cite{li2008time, baer2008mapping}}.

The exchange-correlation potential $ V_{R}^{\mathrm{xc}} $ is obtained from the KS potential by subtracting the external potential $V_{R}^{\mathrm{ext}}(\tau)$ and the Hartree potential $V_{R}^{\mathrm{H}}(\tau)= \sum_{\sigma} n_{R \sigma}(\tau)\, U$. (We show our main results for $ U=2 $ only, as different moderate values of $ U $ give qualitatively similar results.)

In the strongly localized half-filled chains, the ground-state density of both the interacting and noninteracting systems is unity on each site, so that the ground-state $ V_{R}^{\mathrm{Hxc}} $ and $ V_{R}^{\mathrm{xc}} $ are both independent of $ R $. In these systems, for the whole range of time observed, a potential $ V_{R}^{\mathrm{xc}} $ appears that tends to screen the external perturbing field (Fig.~\ref{fig:vxc}), having a gradient with the same sign and order of magnitude as that of the Hartree potential (in strong contrast to the adiabatic LDA which would predict an opposite field). (The instantaneous build-up of our xc electric field, resulting from the application of a `sudden' perturbation, is consistent with the short-range interaction lattice model of Ref.~\onlinecite{schmitteckert2013exact}.) The slope of $ V_{R}^{xc} $, after an initial increase (sites on the left) or decrease (sites on the right) from zero, varies slightly in time and space around the value of the ground-state exchange-correlation potential for the perturbed Hamiltonian $ V_{R}^{\mathrm{pgs-xc}} $ (shown by red dots), giving rise to a ladder-like form.

The generation of an exchange-correlation electric field which counteracts the perturbing potential was previously observed in the response of a periodic insulating solid subjected to an electric field within ground-state DFT~\cite{PhysRevLett.74.4035}. This xc electric field was attributed to the dependence of the exchange-correlation energy on the polarization -- which in effect describes the ultra-long-range part of the density -- implying that DFT functionals, in order to describe global changes in the potential in terms of the electron density, have to be ultra-non-local. The strongly non-local nature of the exchange-correlation potential has also been observed and studied in polymers and other molecular chains using time-dependent current density functional theory (TDCDFT)~\cite{PhysRevLett.88.186401} and optimized-effective-potential (OEP) DFT~\cite{PhysRevLett.93.213002}.  All these studies show that the counteracting xc electric field is not reproduced by simple local-density approximations.

\begin{figure}[h]
\begin{center}
\includegraphics[scale=0.34]{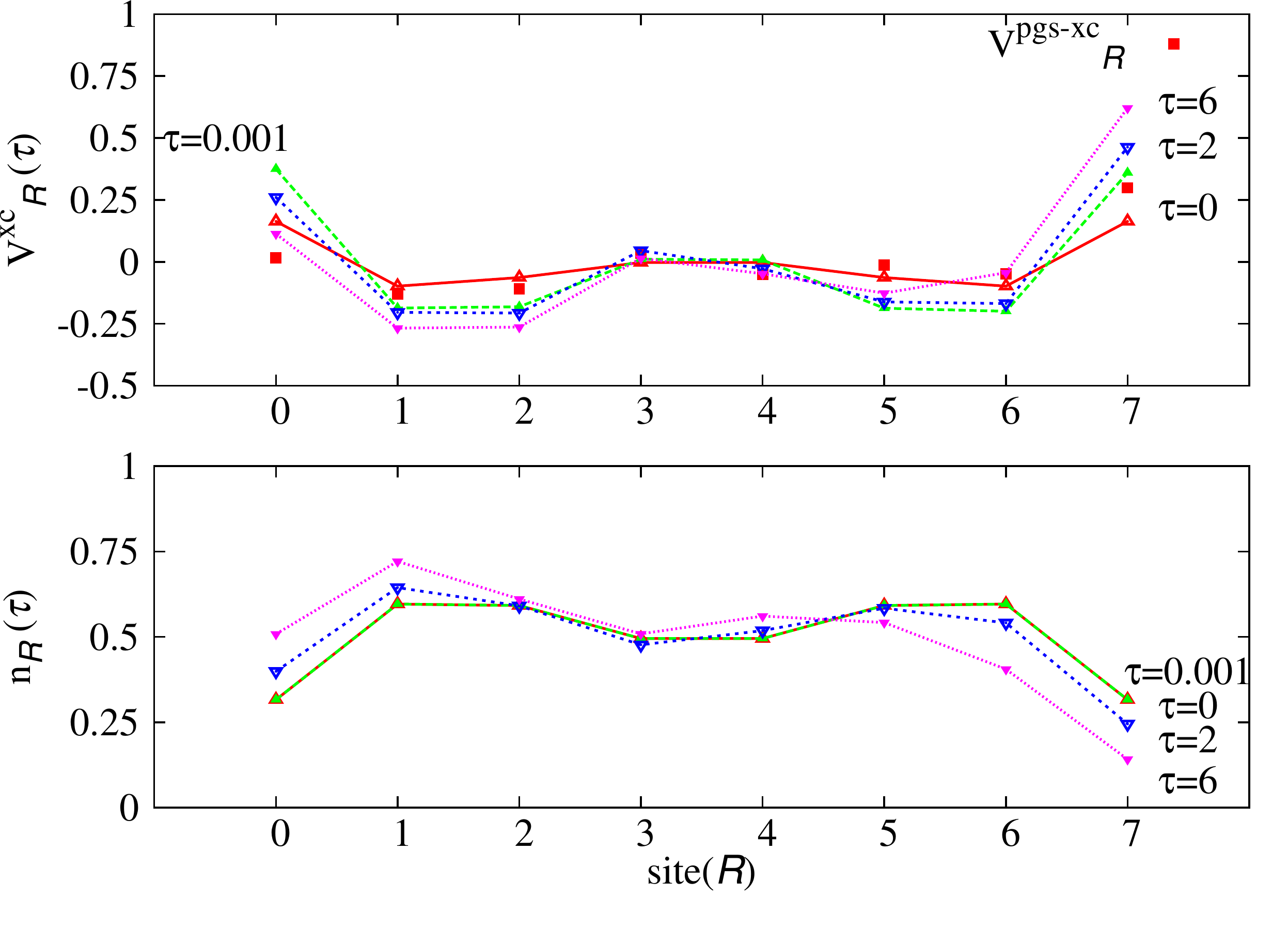}
\caption[Vn2]{(color online) Exchange-correlation potential and charge density, as a function of the position for different times $\tau$, for a quarter-filled 8-site chain subjected to a sudden $ V_{R}^{\mathrm{ext}}(\tau) $. $ V_{R}^{\mathrm{xc}} $ shows a different behavior with respect to the half-filled case: there is no screening xc electric field and the spatial variation is concentrated towards the ends of the chain (more LDA-like behavior). The sharp jump in $ V^{\mathrm{xc}} $ during the first time-step, and its similarity to that of the perturbed ground state, $ V_{R}^{\mathrm{pgs-xc}} $, as observed in the half-filled chains, still hold.}
\label{fig:vxc2}
\end{center}
\end{figure}

Chains in the quarter-filling configuration, or similar configurations for an odd number of sites, show a quite different behavior.  In particular, the ground state $ n_{R} $ depends on the interaction strength $U$, so that the ground state $ V_{R}^{\mathrm{xc}} $ is already non-zero. $ V^{\mathrm{xc}} $ no longer adopts the form of a screening electric field, and its spatial variation is concentrated towards the ends of the chain; its relationship to the local density reflects a more LDA-like behavior (Fig.~\ref{fig:vxc2}). (The `slow' external potential gives qualitatively similar results.) The electric field observed is consistent with the results found by Akande \textit{et al.}~\cite{PhysRevB.82.245114} for 1D Hubbard chains away from half filling in order to study the electric field response.

\section{Adiabatic and instantaneous ground-state $ V^{\mathrm{xc}} $ }
We compare the exact $ V_{R}^{\mathrm{xc}}(\tau) $ with two approximate potentials based on the notion of adiabatic correctness of ground-state DFT for sufficiently slowly varying external potentials: (a) the `adiabatically exact' exchange-correlation potential $ V_{R}^{\mathrm{ad-xc}}(\tau) $ for which the ground-state charge density of the non-interacting system is equal to the actual many-body charge density at each time $ \tau $, and (b) the `instantaneous ground-state' exchange-correlation potential $ V_{R}^{\mathrm{igs-xc}}$ which is simply the exact ground-state exchange-correlation potential associated with the instantaneous \textit{external potential}.  (In the case of the sudden external potential, $ V_{R}^{\mathrm{igs-xc}}$ therefore corresponds to the time-independent perturbed-ground-state potential $ V_{R}^{\mathrm{pgs-xc}}$, which was shown in Figs. ~\ref{fig:vxc} and \ref{fig:vxc2} by red dots.) We apply these two approximations to half- and quarter-filled chains, considering both `sudden' and `slow' external potentials.

An exact adiabatic approach has previously been used for studying various systems (e.g. ~\cite{PhysRevLett.100.153004}), and the `instantaneous ground-state' approach has been applied to Hubbard systems with a localized time-dependent perturbing potential \cite{PhysRevLett.101.166401}.  The performance of ``exact adiabatic'' approximations has also been explored in two recent papers~\cite{fuks2013hubbarddimer,fuks2013arxiv} in the context of resonant charge-transfer excitations in response to an oscillatory electric field.

\begin{figure}[h]
\begin{center}
\includegraphics[scale=0.34]{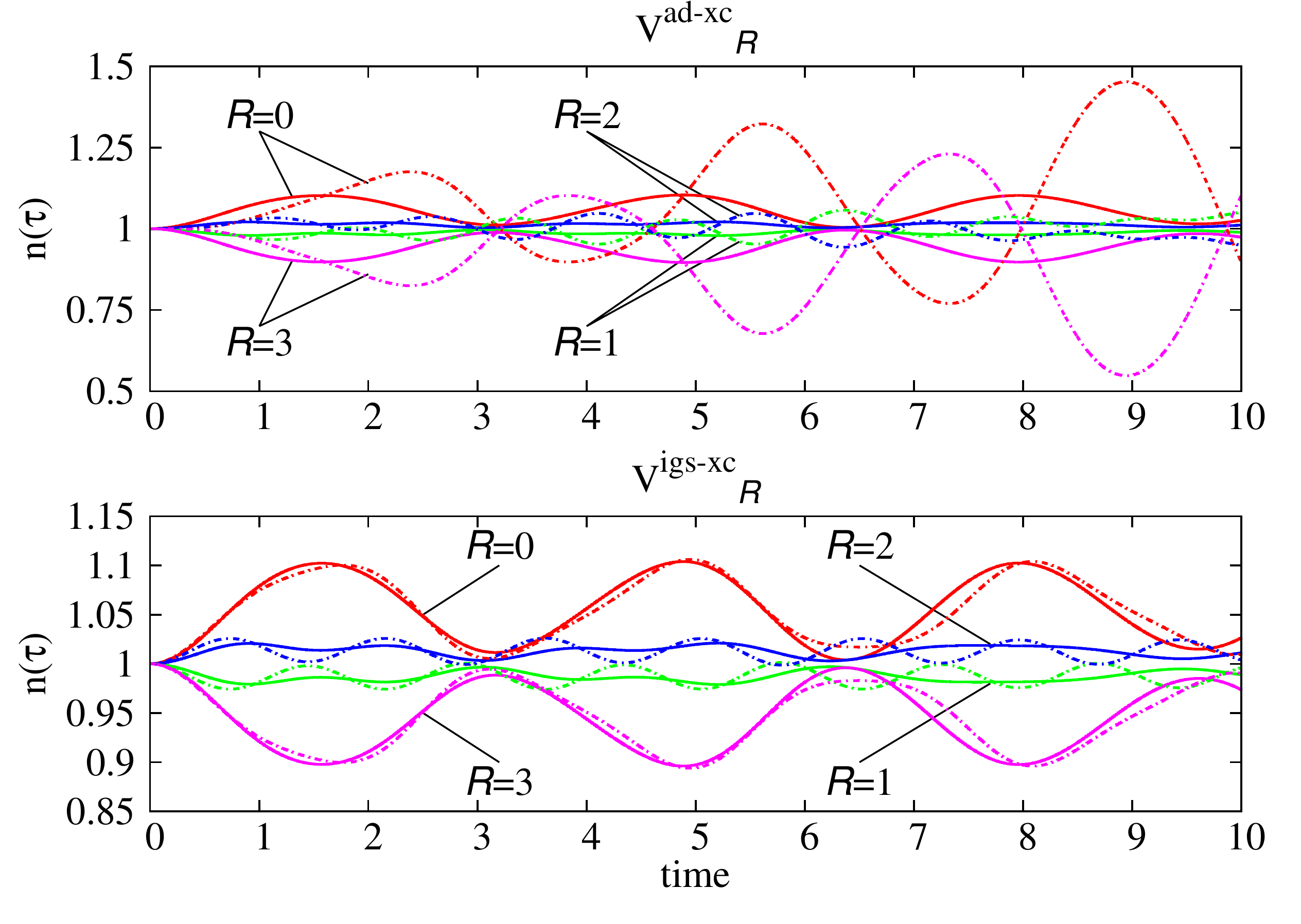}
\caption[AdIst]{(color online) Exact many-body charge density (full lines) and approximated charge density (dashed lines) for a 4-site half-filled chain subjected to the `sudden' $ V_{R}^{\mathrm{ext}}(\tau) $ obtained using the two approximate potentials $ V_{R}^{\mathrm{ad-xc}}(\tau) $ and $ V_{R}^{\mathrm{igs-xc}} $. The latter is a surprisingly good approximation even for a fast applied field, while the former is quite unable to reproduce the exact density.}
\label{fig:AdIst0}
\end{center}
\end{figure}

\begin{figure}[h]
\begin{center}
\includegraphics[scale=0.34]{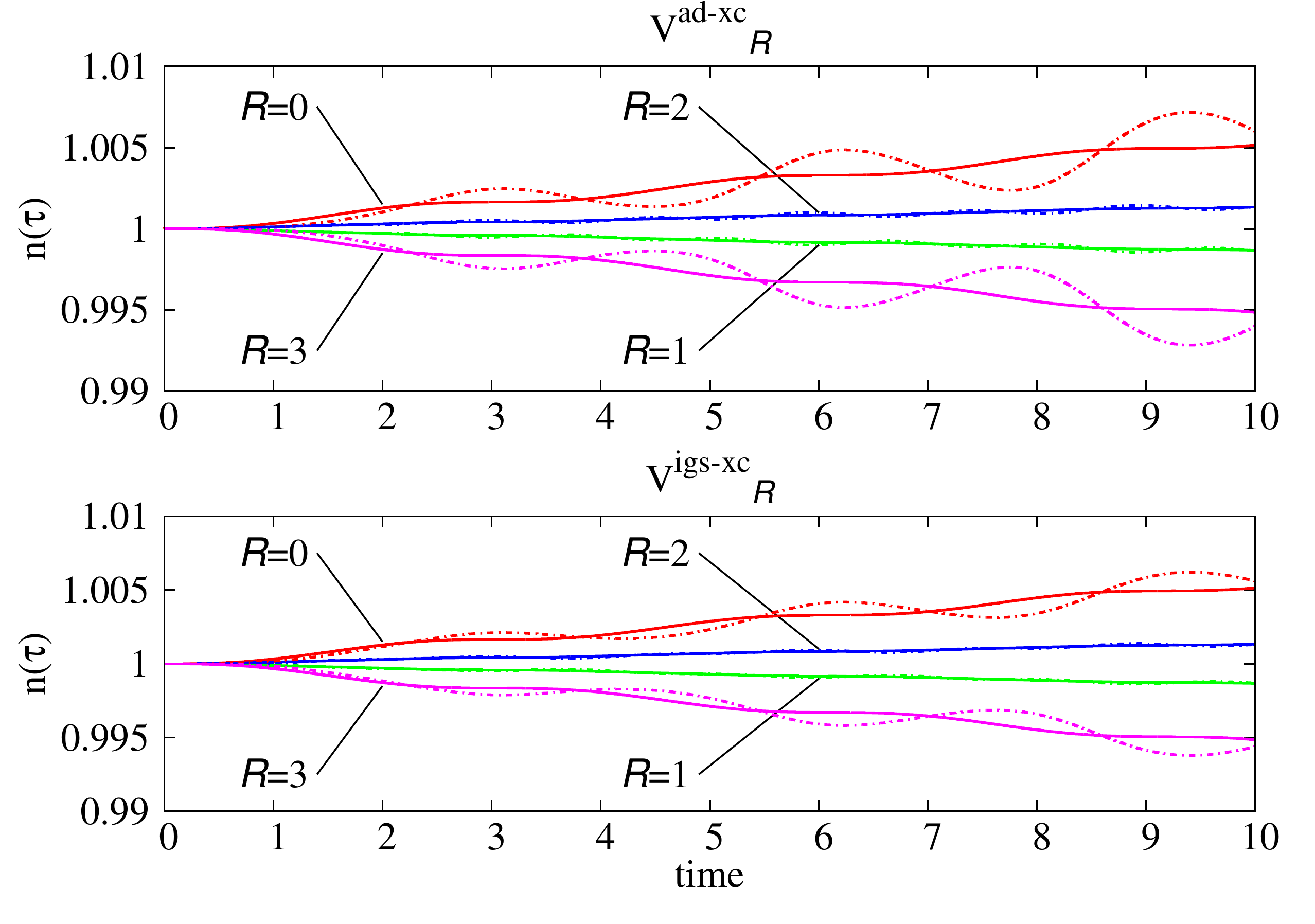}
\caption[AdIst]{(color online) Exact many-body charge density (full lines) and approximated charge density (dashed lines) for a 4-site half-filled chain subjected to a `slow' $ V_{R}^{\mathrm{ext}}(\tau) $ obtained using the two approximated potentials $ V_{R}^{\mathrm{ad-xc}}(\tau) $ and $ V_{R}^{\mathrm{igs-xc}} $. The density reproduced by $ V_{R}^{\mathrm{igs-xc}} $ is closer to the real one than the density reproduced by $ V_{R}^{\mathrm{ad-xc}}(\tau) $.}
\label{fig:AdIst}
\end{center}
\end{figure}

In our half-filled chains, for the `sudden' potential, $ V_{R}^{\mathrm{igs-xc}}(\tau) $ works extremely well in reproducing the exact density, while the adiabatically exact approximation results are strikingly poor (Fig.~\ref{fig:AdIst0}). The fact that the latter does not work well is not an unexpected outcome: an adiabatic approximation is expected to be good only for slowly varying systems, according to the adiabatic theorem \cite{adiabatic.theorem}. For the slow perturbing potential, both $ V_{R}^{\mathrm{ad-xc}}(\tau) $ and $ V_{R}^{\mathrm{igs-xc}}(\tau) $ perform well, once the potential changes sufficiently slowly, with $ V_{R}^{\mathrm{igs-xc}}(\tau) $ once again giving a better approximation than  $ V_{R}^{\mathrm{ad-xc}}(\tau) $ (Fig.~\ref{fig:AdIst}). (For studying the range of applicability of the adiabatic approximation, different velocities of variation have been considered: if the potential is applied slowly enough, the relative difference between the exact and the approximate density is independent of the speed of variation.)  It appears that the character of the strong correlation present in the half-filled Mott-Hubbard-insulating chains is relatively resistant to time-dependent excitation, and remains well described by the \textit{ground} state of the instantaneous external potential.  In contrast, in the less strongly-correlated quarter-filled chains, neither adiabatic approximation is good, even for extremely slowly-varying potentials.

\section{Assessment of local approximations}

In the limit of slow spatial variations in adiabatic potentials, the success of the local density approximation (LDA) in static density-functional theory is extendable to the time-dependent regime (e.g. \cite{Phys.Rev.Lett.45.204}). 
For {\it nonadiabatic} (i.e. frequency-dependent) functionals, it has been proven that approximations that are local or semilocal in the charge density are generally 
impossible\cite{Phys.Rev.Lett.73.2244,Phys.Lett.A.209.206}: this is known as the {\it ultranonlocality} problem. 
However, a local nonadiabatic functional of the {\it current} density is possible \cite{Phys.Rev.Lett.77.2037}, and a uniqueness theorem for a time-dependent 
current-density functional theory has been proven \cite{Phys.Rev.A.38.1149}.

For half-filled Hubbard chains, we saw that, while the usual adiabatic approximation fails, a different choice of adiabatic approximation performs remarkably well, which raises the question of whether a local approximation to the XC potential may be constructed in terms of the charge density alone, or with the current density, or not at all.

Previous investigations of local functionals for Hubbard chains have mainly focused on the Bethe-Ansatz-based LDA (BALDA) approach of Ref. \onlinecite{lima2003BALDA}. Even if nonlocal and memory effects are neglected, it has been shown that adiabatic approximations are able to describe accurately the evolution of systems subjected to TD {\it local} perturbations~\cite{PhysRevB.78.195109}, but to a much lesser extent in the quantum transport regime~\cite{PhysRevB.84.115103}.

In order to study the feasibility of an approximation for the exchange-correlation potential based on the local electronic charge density (in the philosophy of the LDA) and the local current density, we examine the relation between $ V_{R}^{\mathrm{xc}}(\tau) $, $ n_{R}(\tau) $ and $ J_{R}(\tau) $. For this purpose, the on-site current density $ J_{R}(\tau) $ is defined as the average value of the two inter-site current densities $ J_{R \pm 1/2}(\tau) $ surrounding the site (using zero beyond the chain ends). We are also interested in the dependence of the non-equilibrium part of the KS electric field on the local current density, as suggested by Ref.~\onlinecite{PhysRevLett.109.036402} on the basis of calculations for a model semiconductor.
 
\begin{figure}[h]
\begin{center}
\includegraphics[width=0.45\textwidth]{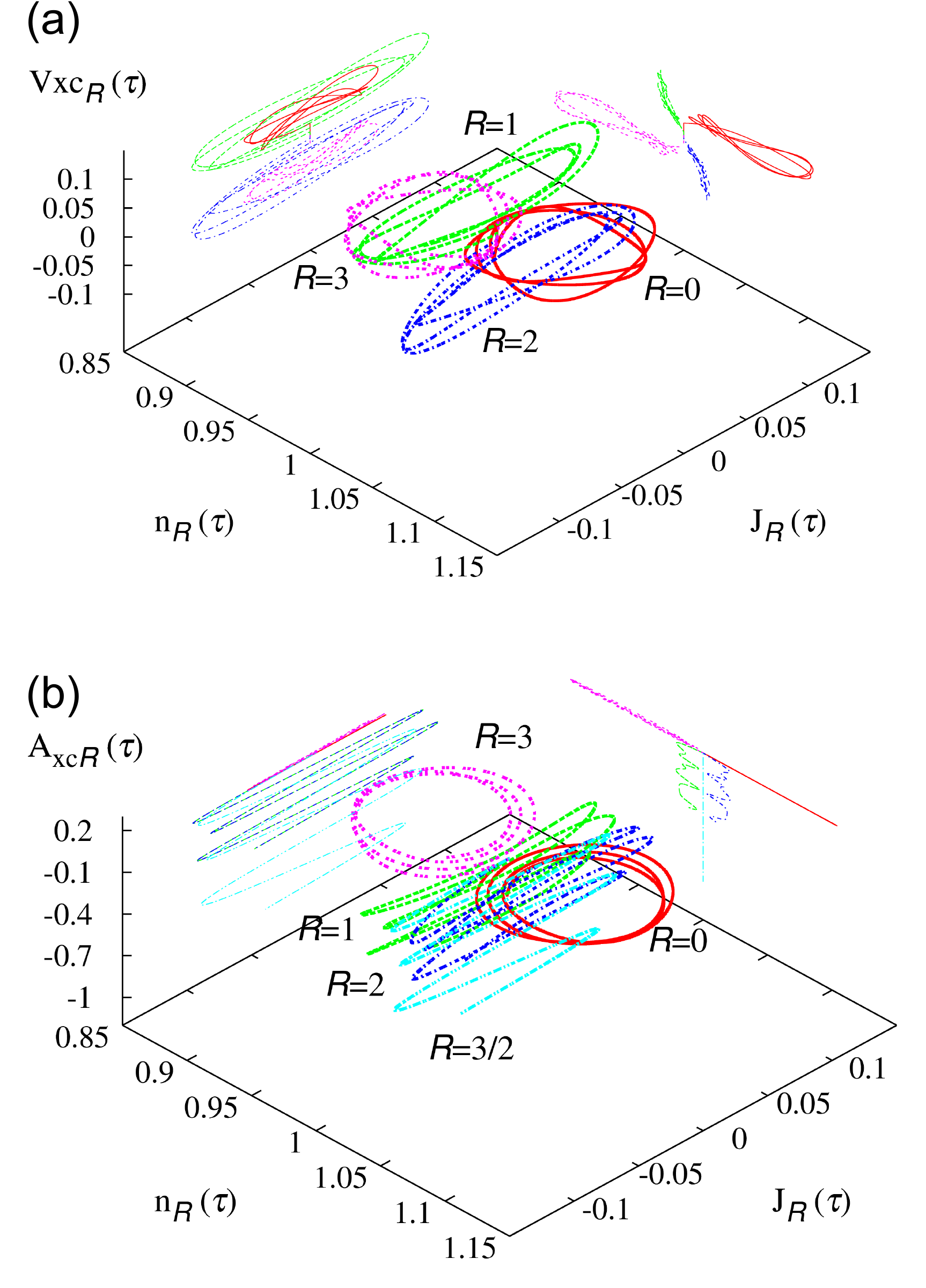}
\caption[n,J,Vxc]{(color online) (a) Exchange-correlation scalar potential $ V_{R}^{\mathrm{xc}}(\tau) $ and (b) the equivalent vector potential $A_{R}^{\mathrm{xc}}(\tau)$ for a 4-site half-filled chain subjected to a sudden perturbing potential, plotted against the local density $ n_{R}(\tau) $ and the on-site current density $ J_{R}(\tau) $. For adiabatic approximations of $ V^{\mathrm{xc}} $ and $ A^{\mathrm{xc}} $ based on the local values of $ n_{R} $ and $ J_{R} $ to hold, it would have to be possible to plot the potentials on universal surfaces (the same for all the sites) with respect to one or both of the two quantities. The projections onto the $ V^{\mathrm{xc}} $-$ J $, $ V^{\mathrm{xc}} $-$ n $, $ A^{\mathrm{xc}} $-$ J $ and $ A^{\mathrm{xc}} $-$ n $ back-planes are also shown; these serve to clarify the spatial form of the trajectories, and also directly illustrate the limitations of local approximations based on $n$ or $J$ only.}
\label{fig:nJvxc}
\end{center}
\end{figure}

We first consider the dependence of $V^{\mathrm{xc}}$ on the local $n$ and $J$.  In Fig.~\ref{fig:nJvxc}(a) we show the relation between $ V^{\mathrm{xc}}(\tau) $, $n_{R}(\tau)$ and $ J_{R}(\tau) $ for the four-site half-filled chain subjected to the sudden $ V^{\mathrm{ext}}(\tau) $ (in this configuration the ground-state $ V^{\mathrm{xc}}$ is zero, so that the plotted quantity coincides with the dynamical part of the potential). Even considering that the exchange-correlation potential is defined up to an additive TD constant, there is clearly no unique relationship between the potential and the local charge density, the local current density, or both together.
Specifically, if a dependence of $V^{\mathrm{xc}}$ on the local densities existed, it would have to be possible to plot the potentials for the various sites on a universal surface with respect to $n_{R}$ and $ J_{R}$. In Fig.~\ref{fig:nJvxc} we fix the TD constant so that $V^{\mathrm{Hxc}}$ is zero on $R=0$ for each $\tau$. In this case, the potentials on $R=0$ and $R=3$ varies in a small range during the whole time observed (this can be easily seen with the help of the projections onto the $ V^{\mathrm{xc}} $-$ n $ plane). One can envisage moving these two trajectories onto essentially a single surface by appropriate modification of the TD constant. However, comparing the potentials for $R=0$ and $R=3$ with those on the other two sites, it is easy to see that the latter will never lie on the same surface.   

Another argument against the feasibility of a local approximation can be found looking at $ V^{\mathrm{xc}}$ in the early time-steps (Fig.~\ref{fig:vxc}). In half-filled chains the ground-state KS potential is the same on each site. After the first time-step, a sudden change in $ V^{\mathrm{xc}}$ occurs which is not simply due to the discretization $ \Delta\tau $ and which is not reflected locally in charge and current densities, and thus not reproducible in a local density approximation. Even in quarter-filled chains, where the exchange-correlation potential shows a more LDA-like behavior, it is still not possible to find an universal surface.

Also shown in Fig.~\ref{fig:nJvxc}(b) is the vector potential
\begin{equation}
 A_{R}^{\mathrm{xc}}(\tau) = \int_{0}^\tau~d\tau'~\frac{\partial V_{R}^{\mathbf{xc}}(\tau')}{\partial x} \nonumber
\end{equation}
as an instantaneous function of the local charge and current densities.  This quantity is introduced by an alternative choice of gauge -- the `velocity' gauge -- in which the time-dependence previously in $V^{\mathbf{xc}}(\tau)$ is moved into the vector potential $A^{\mathbf{xc}}(\tau)$. Once again, if the instantaneous vector potential were approximately given by a local functional of the charge and/or current density, $A_{R}^{\mathrm{xc}}(n, j)$ should lie on a universal surface. As we can see, this is not the case, so the vector potential is not uniquely determined by the local densities, either individually or together.

We have also explored the use of direct non-linear optimization techniques to find the best universal local approximation, based on approximating $ V_{R}^{\mathrm{xc}} $ as the sum of two parts: one based on the local density and the other whose \textit{gradient} is based on the local current.  No such form proves to be capable of yielding a satisfactory representation of $ V_{R}^{\mathrm{xc}} $ for the sites within one system, let alone a range of systems.  

Thus we see that, even where an adiabatic approximation to the time-dependent XC potential is possible, neither the scalar potential nor its gradient are uniquely determined by the local charge and current density, and that the nonlocality in the potential survives the reformulation of the time-dependent density-functional theory in terms of the current density and vector potential.
    
The vector potential does provide some insight into the exchange-correlation functional that the scalar potential does not. Its functional dependence on the current density is observed to be the same for the $R=1$ and $R=2$ sites, while the charge-density-dependence switches sign, reflecting the underlying spatial antisymmetry of the time-dependent density. At the $R=3/2$ site, the time-derivative of the charge density is also zero due to symmetry, and thus we see the functional dependence of the vector potential on the current density alone.

However, the vector potential as a functional of $J$ appears to grow approximately linearly with time, i.e. is {\it ultranonlocal} in time, thus in transforming the gauge we have simply traded one ultranonlocality for another. One can see how this is necessarily the case by considering components of the exchange-correlation {\it electric field}, such as part of the xc electric field switched on immediately after $\tau=0$, that are approximately constant with time. Implementing these in the velocity gauge yields a vector potential with components that are linear in time.

As such, purely scalar or purely vector local potential functionals would fail to capture the correct nonequilibrium physics. Rather, it appears more useful to cast future functionals in terms of both scalar and vector exchange-correlation components together with, if necessary, electric field components (as discussed in Ref. \onlinecite{PhysRevLett.109.036402}), to account for nonlocal dependence.

\section{Time-dependence of the correlation}

The pair correlation function (PCF) provides direct insight into the nature of the electronic correlation. The PCF can be written
\begin{equation*}
\Gamma _{\sigma , \sigma '}(R,R')= \frac{\langle \psi \vert \widehat{\Psi}_{R\sigma} ^\dagger \widehat{\Psi}_{R'\sigma '} ^\dagger \widehat{\Psi}_{R'\sigma '} \widehat{\Psi}_{R\sigma}  \vert \psi \rangle}{n_{R,\sigma} n_{R',\sigma '}}
\end{equation*}
(if $ R=R'$ and $ \sigma = \sigma ' $, $ \Gamma = 0 $). We consider the on-site PCF ($ \uparrow \downarrow $). In this case, if $ U=0 $, the PCF is equal to one for each $ R $. Increasing the interaction, its ground-state value decreases towards zero.

\begin{figure}[h]
\begin{center}
\includegraphics[scale=0.34]{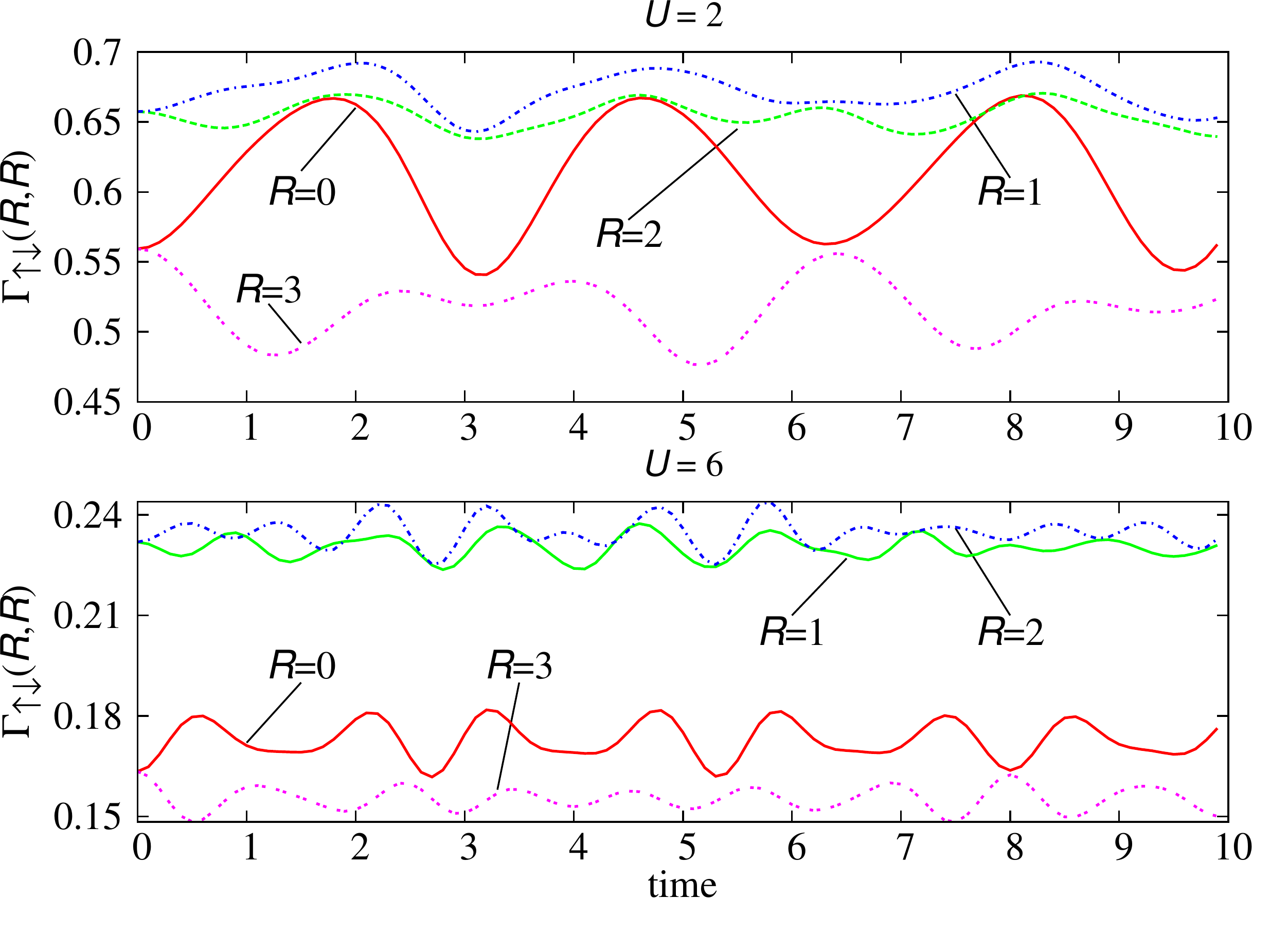}
\caption[Pcf]{(color online) On-site pair correlation function as a function of time for a 4-site half-filled chain. Passing from $ U=2 $ to $ U=6 $, the ground state $ \Gamma _{\uparrow , \downarrow}(R,R) $ decreases and the variations of the PCF during the time evolution with respect to the ground state value are smaller.}
\label{fig:Pcf}
\end{center}
\end{figure}

In general, if one considers half-filled chains, the most striking features of $ \Gamma _{\sigma , \sigma '}(R,R) $ is that its value varies only slightly during the time evolution with respect to the ground-state value for higher magnitudes of $ U $: the nature of the electronic correlation is, to a good approximation, frozen in its ground-state form (Fig.~\ref{fig:Pcf}). This feature seems to be consistent with the good results obtained using the instantaneous-ground-state potential $ V_{R}^{\mathrm{igs-xc}}(\tau) $ as an approximate KS functional. On the other hand, quarter-filled chains shows quite a different behavior: in this case, the nature of the correlation shows a stronger time dependence, as does the electron density $n$. 

\section{Conclusion}

In conclusion, we have shown that in half-filled Hubbard chains the time-dependent electronic correlation remains dominated (especially for larger $ U $) by the strongly localized ground state: adiabatic approximations for the exchange-correlation potential based on the instantaneous ground state of the external potential perform remarkably well even when exact adiabatic charge-density-based approximations fail. This feature is consistent with the computed form of the pair correlation function. On the other hand, for quarter-filled chains, due to stronger time dependence of both the density itself, and the changing nature of the correlation, the exact exchange-correlation potential acquires time-dependent features that are absent from both types of adiabatic functional. In all systems studied the exact exchange-correlation scalar potential has a spatial form that has a strongly non-local dependence on the charge (and current) densities, while the corresponding vector potential is temporally nonlocal.  Our results indicate that the search for accurate approximate functionals for time-dependent DFT should focus on functionals that build in spatial and temporal non-locality in the densities, for example through mixed scalar/vector xc-potentials, or explicit dependence on the Kohn-Sham wavefunctions.

\bibliographystyle{apsrev4-1} 
\bibliography{Bibtex}

\end{document}